\documentclass[aps,prl,twocolumn,showpacs,preprintnumbers,amsmath,amssymb]{revtex4}

\usepackage{graphicx}
\usepackage{dcolumn}
\usepackage{bm}

\begin{document}


\title{Validity of the additivity principle in the weakly asymmetric exclusion processes with open boundaries}
\author{Mieke Gorissen$^1$}
\author{Carlo Vanderzande$^{1,2}$}
\affiliation{%
$^1$Faculty of Sciences, Hasselt University, 3590 Diepenbeek, Belgium. \\
$^2$Instituut Theoretische Fysica, KULeuven, 3001 Heverlee, Belgium.
} 


\date{\today}

\begin{abstract}
The additivity principle allows a calculation of current fluctuations and associated density profiles in large diffusive systems. In order to test its validity in the weakly asymmetric exclusion process with open boundaries, we use
a numerical approach based on
the density matrix renormalisation. With this technique, we determine the cumulant generating function of the current and the density profile corresponding to atypical currents in finite systems. We find that these converge to those predicted by the additivity principle. No evidence for dynamical phase transitions is found. \end{abstract}

\pacs{05.40.-a, 02.50.-r, 05.70.Ln, 44.10.+i}
\maketitle
Systems driven out of equilibrium by putting them in contact with reservoirs at different chemical potential or temperature develop currents. It is a main problem of non equilibrium statistical mechanics to determine how the average and fluctuations of these currents can be derived from microscopic dynamics. In recent years, considerable progress has been made in this direction. Firstly, it was found that current fluctuations have symmetries as expressed in the Gallavotti-Cohen \cite{Gallavotti95,Lebowitz99} theorem. These symmetries are macroscopic manifestations of microscopic time reversibility. Secondly, Bodineau and Derrida formulated an additivity principle (AP) that allows one to calculate the whole distribution of current fluctuations once the first two cumulants are known \cite{Bodineau04}. The AP should hold for one-dimensional, diffusive systems  and was validated in the symmetric exclusion processes \cite{Derrida07} and the Kipnis-Marchioro-Presutti (KMP) model of heat conduction \cite{Hurtado09}. Most recently, it was also found to hold in three-dimensional {\it deterministic} models of heat conduction \cite{Saito11}. Independently, Bertini {\it et al.} developped a large deviation theory for density and current fluctuations in stochastic lattice gases \cite{Bertini01,Bertini05}. The predictions of this Hydrodynamic Fluctuation Theory coincide with those of the AP when the fluctuations are time-independent. Interestingly, it was found that for sufficiently large fluctuations a dynamical phase transition can occur to a phase where density and current fluctuations become time-dependent \cite{Bertini05}. This transition was observed in a weakly asymmetric exclusion process on a ring \cite{Bodineau05} and more recently in the KMP model \cite{Hurtado11}, also on a ring.  This type of dynamical transitions can occur in situations that are not allowed in equilibrium and have been conjectured to be of relevance to such issues as breaking of chiral or CP-symmetry \cite{JonaLasinio10}. 

In this Letter, we study the current fluctuations in the weakly asymmetric exclusion process (WASEP) with open boundaries. We calculate the cumulant generating function of the current and the density profile giving rise to an atypical current using the AP. This extends earlier work \cite{Bodineau06}. We compare these results with those coming from calculations in finite systems using the density matrix renormalisation group (DMRG). We recently showed how that approach, first introduced to study low temperature properties of quantum systems \cite{White92}, can be applied to determine the cumulant generating function of the current or the activity of stochastic systems \cite{Gorissen09}. In the present work,  we are the first to calculate density profiles corresponding to current fluctuations with the DMRG. 
We find that for sufficiently large systems our results converge to those predicted by the AP in the whole range of parameters investigated, further validating this principle. We find no evidence for a dynamical transition in this case.

{\it Model $-$} In the asymmetric exclusion process (ASEP) \cite{Derrida07}, each site $i$ of a lattice of size $N$ can be empty or occupied by one particle. The dynamics is that of a Markov chain where particles  jump to the right or left with different rates ($p$ resp. $q$). This describes the effect of an external field $E=\ln(p/q)$. In this Letter we will discuss the case where $p=1+\nu/(2N), q=1-\nu/(2N)$ referred to as the WASEP. This is a diffusive model where the AP should be applicable. We will consider the case of open boundaries where at its left (right) side the system is in contact with a reservoir at density $\rho_a$ ($\rho_b$). We will assume $\rho_a > \rho_b$ and $\nu>0$. The model will evolve to a non equilibrium steady state (NESS). 

{\it Current fluctuations from the additivity principle $-$} We are interested in the total number of particles $Q_T$ passing in a large time $T$ through the system. For a large system, and using a continuum description in terms of $x=i/N \in [0,1]$, the average current equals
\begin{eqnarray}
\frac{\langle Q_T\rangle}{T} = - \frac{D(\rho)}{N} \frac{\partial \rho}{\partial x} + \frac{\nu \sigma(\rho)}{N}\equiv \frac{j^*}{N}
\label{0}
\end{eqnarray}
The first term is Fick's law and the second is the current due to the field $E$ in linear response. 
For the WASEP, the diffusivity $D(\rho)=1/2$ and the mobility $\sigma(\rho)=\rho(1-\rho)$ \cite{Spohn91} where $\rho(x)$ is the particle density. In this Letter, we are interested in the fluctuations of the current around the average value $j^\star$. 
For $N$ very large, the probability $P_N$ to observe an integrated current $Q_T=jT/N$ has the form
\begin{eqnarray}
P(\frac{Q_T}{T}=\frac{j}{N},\rho_a,\rho_b) \sim \exp\left[ \frac{T}{N} G(j,\rho_a,\rho_b)\right]
\label{1}
\end{eqnarray}
The large deviation function $G$ is zero at the average current $j^*$ of the NESS, and is strictly negative for other $j$-values.  According to the AP \cite{Bodineau04}, $G$ can be found from a variational principle
\begin{eqnarray}
G(j,\rho_a,\rho_b) =- \min_{\rho(x)} \left[\int_0^1 \frac{[j - \nu \sigma(\rho(x)) + \frac{1}{2} d\rho/dx]^2}{2\sigma(\rho(x))} dx \right]
\label{2}
\end{eqnarray}
which leads to a Euler-Lagrange equation for $\rho(x)$
\begin{eqnarray}
\left(\frac{1}{2} \frac{d\rho}{dx}\right)^2= (j-\nu \sigma(\rho))^2 + 2 K \sigma(\rho)
\label{3}
\end{eqnarray}
Solution of this equation for given $j$ and subject to the boundary conditions $\rho(0)=\rho_a,\ \rho(1)=\rho_b$ determines the integration constant $K$ and the density profile. For a monotonically decreasing profile, one obtains
\begin{eqnarray}
\int_{\rho_a}^\rho \frac{d\rho}{\left[(j-\nu\sigma(\rho))^2+2K\sigma(\rho)\right]^{1/2}} = - 2 x
\label{4}
\end{eqnarray}
More complicated expressions can be determined for the case that the profile has an extremum. Inserting this solution in (\ref{2}) then gives for the large deviation function
\begin{eqnarray}
G(j,\rho_a,\rho_b) &=& \int_{\rho_a}^{\rho_b} \Big[ \frac{(j-\nu\sigma(\rho))^2 + K \sigma(\rho)}{[(j-\nu \sigma(\rho))^2+2 K \sigma(\rho)]^{1/2}} \nonumber \\ &-&(j-\nu \sigma(\rho))\Big]\frac{d\rho}{2\sigma(\rho)}
\label{4a}
\end{eqnarray}

Instead of (\ref{1}), one can also describe the current fluctuations using the cumulant generating function
\begin{eqnarray}
\mu(s,\rho_a,\rho_b) = \lim_{T \to \infty} \frac{1}{T} \ln \langle e^{s Q_T} \rangle
\label{5}
\end{eqnarray}
which is related to $G$ through a Legendre transform
\begin{eqnarray}
\mu(s,\rho_a,\rho_b)= \frac{1}{N} \max _j \left[ s j + G(j,\rho_a,\rho_b)\right]\equiv \frac{M(s,\rho_a,\rho_b)}{N}
\label{6}
\end{eqnarray}
Inserting (\ref{4a})  gives the cumulant generating function (CGF) in parametric form
\begin{eqnarray}
M(s,\rho_a,\rho_b)&=&\frac{1}{2} \int_{\rho_b}^{\rho_a} \frac{(j-\nu \sigma(\rho))\nu - K}{\left[(j-\nu \sigma(\rho))^2+2K\sigma(\rho)\right]^{1/2}}d\rho \nonumber \\ &+& \frac{\nu}{2}(\rho_b-\rho_a) \label{7}
\end{eqnarray}
and
\begin{eqnarray}
s=\int_{\rho_b}^{\rho_a} \left[\frac{j-\nu \sigma(\rho)}{\left[(j-\nu \sigma(\rho))^2+2K\sigma(\rho)\right]^{1/2}}-1\right]\frac{d\rho}{2\sigma(\rho)}
\label{8}
\end{eqnarray}
This is again the result for a monotonically decreasing profile. The more complicated expressions for a profile with an extremum will be given elsewhere. For given values of $\nu,\rho_a, \rho_b$ and $j$ we have determined the density profile, the large deviation function and the CGF by numerical evalution of the integrals in (\ref{4}), (\ref{4a}), (\ref{7}) and (\ref{8}). 

{\it DMRG - approach $-$} We now want to determine the density profile and the CGF for finite systems to see whether for large $N$ they converge to those predicted by the AP. With standard simulation techniques it is difficult to generate atypical currents since they occur with exponentially small probability. A method to overcome this problem has been proposed in \cite{Giardina06,Tailleur07}.  Yet, this technique becomes less accurate for large fluctuations \cite{Hurtado09} due to statistical errors.
Recently, we proposed a new approach to current fluctuations based on the DMRG, the ideas behind which we now briefly explain \cite{Gorissen11}.

The probability $P({\cal C},t)$ to observe the exclusion process in a given microscopic configuration ${\cal C}$ evolves according to the master equation $\partial_t P({\cal C},t) = \sum_{{\cal C'}}H({\cal C},{\cal C'}) P({\cal C'},t)$ where $H$ is the generator of the process. 
 It is by now well established \cite{Derrida07} that the CGF (\ref{5}) can be obtained from a modified generator $H_s$, which is constructed from $H$ as follows. Let $\alpha({\cal C},{\cal C'})$ be $+1$ ($-1$) when in the transition from ${\cal C}$ to ${\cal C'}$ a particle enters (leaves) the system on its left side. Otherwise $\alpha({\cal C},{\cal C'})=0$. For the off-diagonal elements of $H_s$ one has $H_s({\cal C},{\cal C'})=H({\cal C},{\cal C'})e^{s\alpha({\cal C},{\cal C'})}$ while the diagonal elements of $H$ and $H_s$ are equal. The CGF for a system of $N$ sites, $\mu(s,\rho_a,\rho_b,N)$, then equals the largest eigenvalues of $H_s$. Moreover, let $|R_0\rangle$ and $\langle L_0|$ be the associated right and left eigenvector. 
Consider a dynamical variable $b({\cal C}(t))$ (like the density at a given site) which depends on the microscopic configuration in which the system is at time $t$. It can be shown that the current weighted time-average of $b$, defined as,
\begin{eqnarray}
\langle b(\tau) \rangle_s \equiv \frac{1}{T} \frac{\langle \int_0^T b({\cal C}(\tau)) e^{sQ_\tau} d\tau\rangle}{\langle e^{sQ_T}\rangle} 
\label{9}
\end{eqnarray}
for $T$ large equals $\langle L_0|\hat{b}|R_0\rangle$ 
where $\hat{b}$ is the operator associated to the variable $b$ \cite{Garrahan07}. On the other hand,  the average of $b$ at a large time $T$ defined as
\begin{eqnarray}
\langle b(T)\rangle_s \equiv \langle b({\cal C}(T)) e^{sQ_T}\rangle/\langle e^{sQ_T}\rangle
\label{9a}
\end{eqnarray}
equals $\langle 0|\hat{b}| R_0\rangle$ where $\langle 0|$ is the projection state $\sum_{{\cal C}} \langle {\cal C} |$. So both the CGF and the density profiles can be determined from the largest eigenvalue of $H(s)$, its eigenvectors and the projection state. From a mathematical point of view, solving this problem is similar to that of determining the ground state and its eigenvector for a quantum spin chain, where the main difference is that in the stochastic problem the generator $H_s$ is not Hermitian. One of the most precise approaches to determine ground state properties of quantum chains is the DMRG \cite{White92, Schollwock05}. 
We recently showed that this method also works  well for generalized generators associated with current fluctuations \cite{Gorissen09}. 
Here we give for the first time results on time-averaged density profiles. Since the projection state plays no role in quantum mechanical problems, we had to adapt the DMRG approach in order to also calculate late-time averages. Details of this will be given elsewhere. We can typically obtain reliable results up to $N \approx 120$. As a check of the DMRG approach we have calculated density profiles for the totally asymmetric exclusion process ($p=1, q=0$) at $s=0$ where exact results for finite $N$ exist \cite{Derrida93} and have found perfect agreement. We are not aware of any exact results for the density of the ASEP in finite systems and for $s\neq0$.

\begin{figure}
\includegraphics[width=8.0cm]{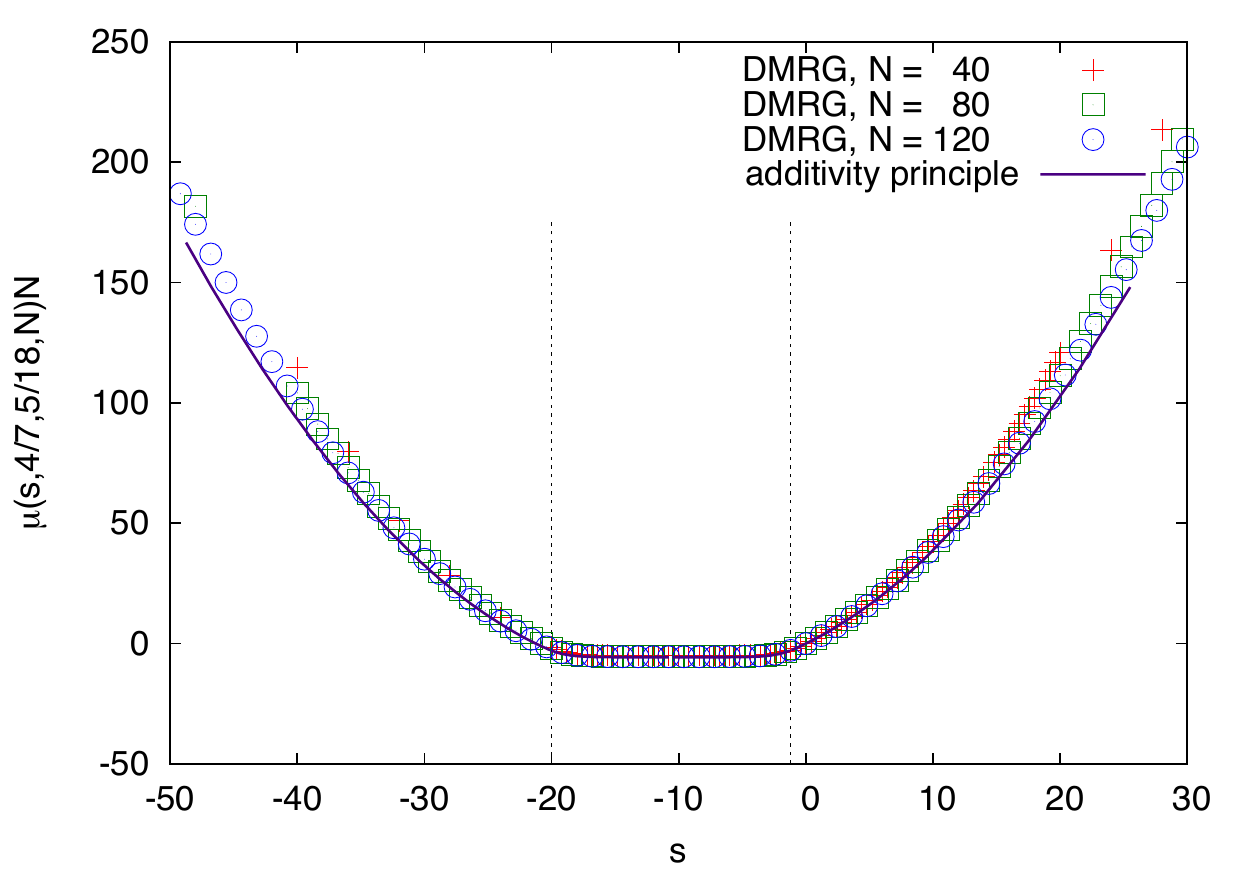}
\caption{\label{fig1} (Color online) Cumulant generating for the WASEP for $\nu=10,\rho_a=4/7, \rho_b=5/18$ from the additivity principle and from the DMRG.}
\end{figure}

{\it Results $-$} We have performed most of our calculations for $\nu=10, \rho_a=4/7$ and $\rho_b=5/18$. In Fig.\ref{fig1} we present our results for the CGF. The full line is the prediction form the AP. In the regime between the vertical dotted lines, the optimal profile has a minimum, otherwise it is monotonically decreasing. The various symbols are DMRG results for $N \mu(s,\rho_a,\rho_b,N)$ at different $N$-values. As can be seen, for increasing system size,  the AP and DMRG results coincide within numerical accuracy in an increasing range of $s$-values. Within the whole $s$-region investigated our numerical data satisfy the Gallavotti-Cohen symmetry $\mu(s,\rho_a,\rho_b,N)=\mu(\Delta-s,\rho_a,\rho_b,N)$ where
\begin{eqnarray}
\Delta=-\ln\frac{(1-\rho_a)\rho_b}{\rho_a(1-\rho_b)} - (N-1)\ln\frac{1-\nu/2N}{1+\nu/2N}
\label{10}
\end{eqnarray}
A dynamical phase transition should show up as a point where the CGF becomes non-analytical. On the scale of Fig. \ref{fig1} this seems to occur where the profile changes from monotonic to one with a minimum. A detailed investigation of the first and second derivative of the CGF near these points  however shows no evidence for non-analyticity. 

\begin{figure}
\includegraphics[width=8.0cm]{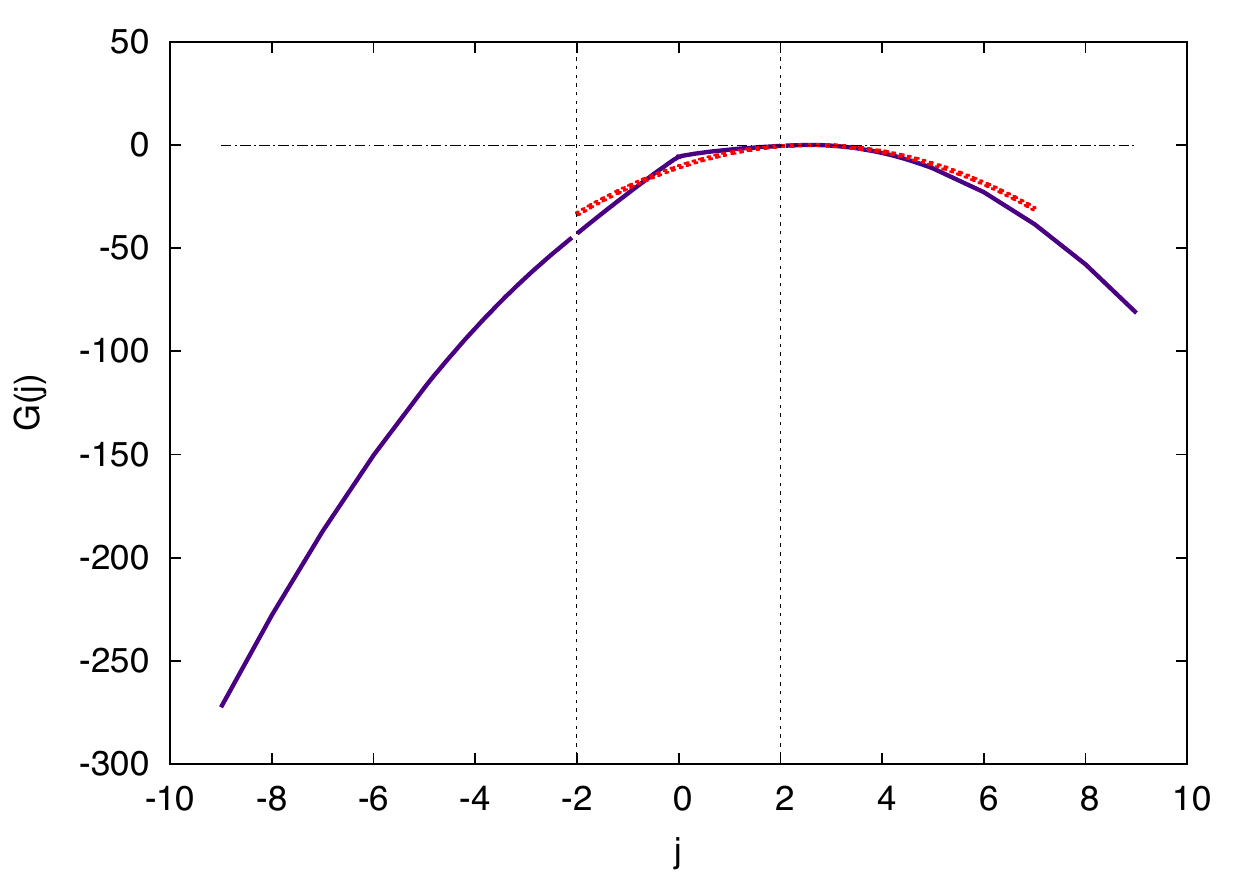}
\caption{\label{fig2} (Color online) Current large deviation function for the WASEP for $\nu=10,\rho_a=4/7, \rho_b=5/18$ from the additivity principle (full line). The dotted line indicates a quadratic fit near the maximum.}
\end{figure}

In Fig. \ref{fig2} we show the large deviation function $G(j)$ as calculated from the AP. This quantity cannot be determined from the DMRG. For small deviations from the average current $j^\star (=2.5845$, as follows from (\ref{1}) and the chosen boundary values), we expect $P(j/N)$ to be Gaussian, so that the LDF is quadratic (dotted line). Clearly, sufficiently large fluctuations are non-Gaussian.

\begin{figure}
\includegraphics[width=8.0cm]{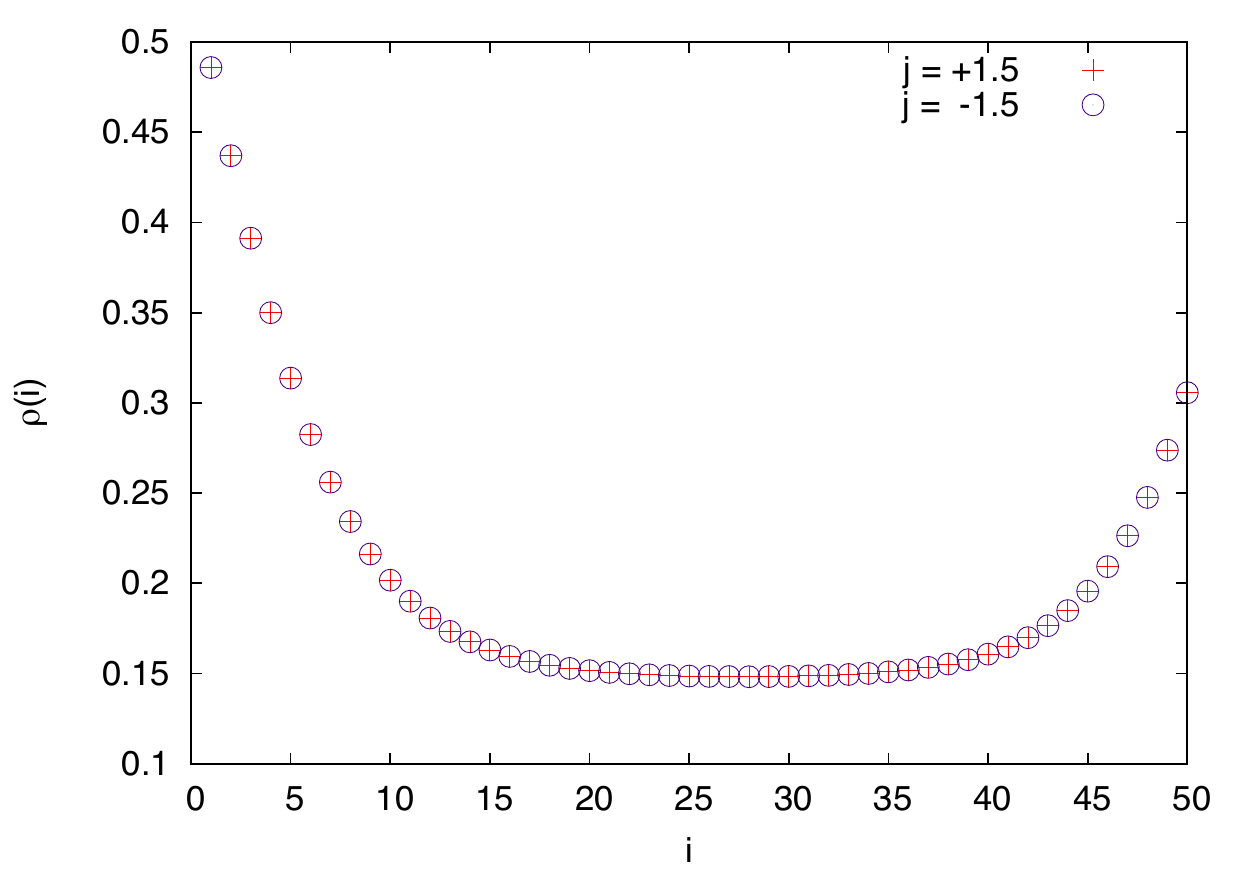}
\caption{\label{fig3} (Color online) Time-averaged density profile in a finite system $N=50$ with current $j=\pm 1.5$ as calculated with the DMRG.}
\end{figure}

We now turn to the density profiles. Firstly, we observe that the time-averaged density profiles are invariant for the transformation $j \leftrightarrow -j$. This is a consequence of time-reversibility, and is a special case of a more general result for isometric current fluctuations that holds in higher dimensions \cite{HurtadoPNAS11}. The invariance of the profiles follows directly from the equations of the AP, but is also valid for finite systems. Fig. \ref{fig3} shows a result for a system with $N=50$ and $j=\pm 1.5$. Density profiles calculated after a large time $T$ do not obey this symmetry. 

\begin{figure}
\includegraphics[width=8.0cm]{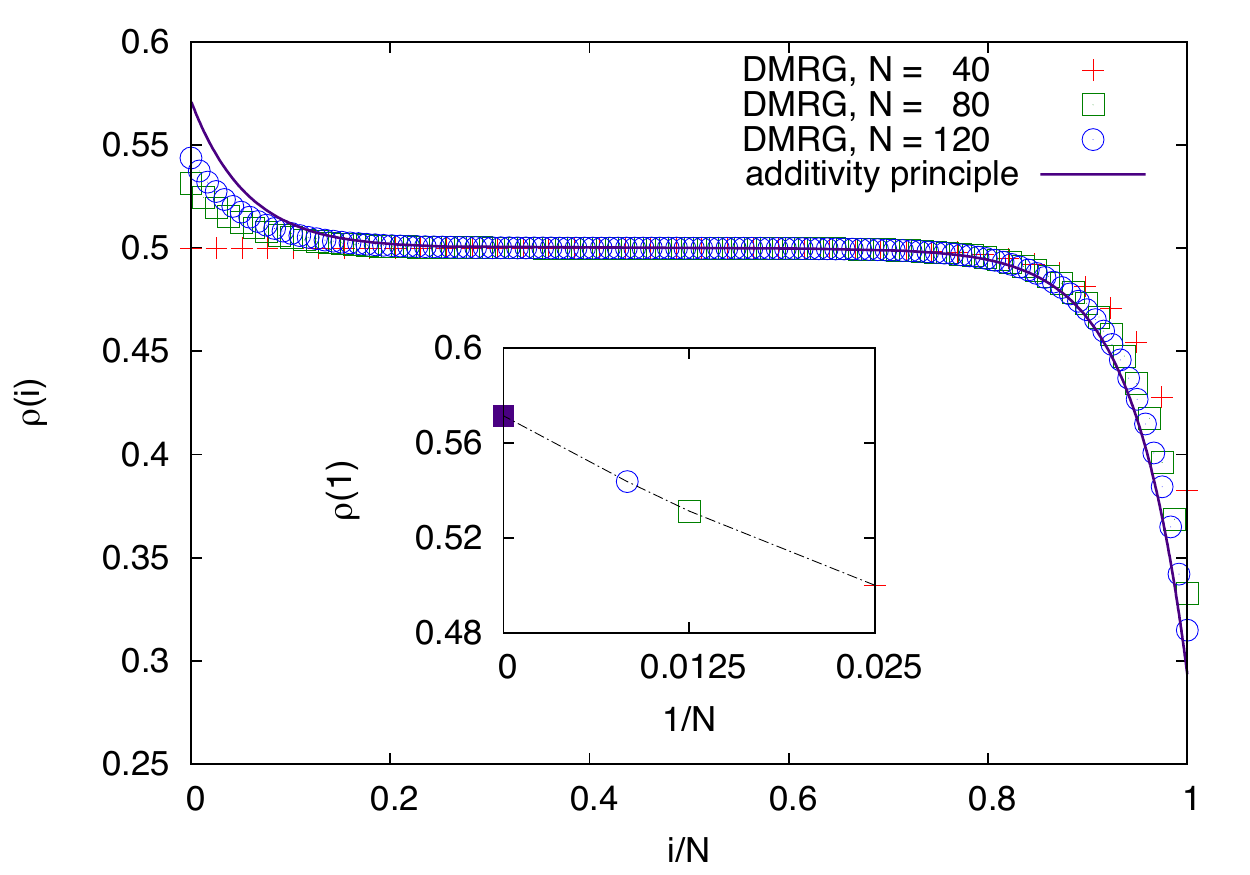}
\caption{\label{fig4} (Color online) Time-averaged density profile at $s=10$ as calculated from the AP (full line) and from the DMRG. The inset shows the density at the first site as a function of $1/N$ together with the prediction of the AP (full square). }
\end{figure}

\begin{figure}
\includegraphics[width=8.0cm]{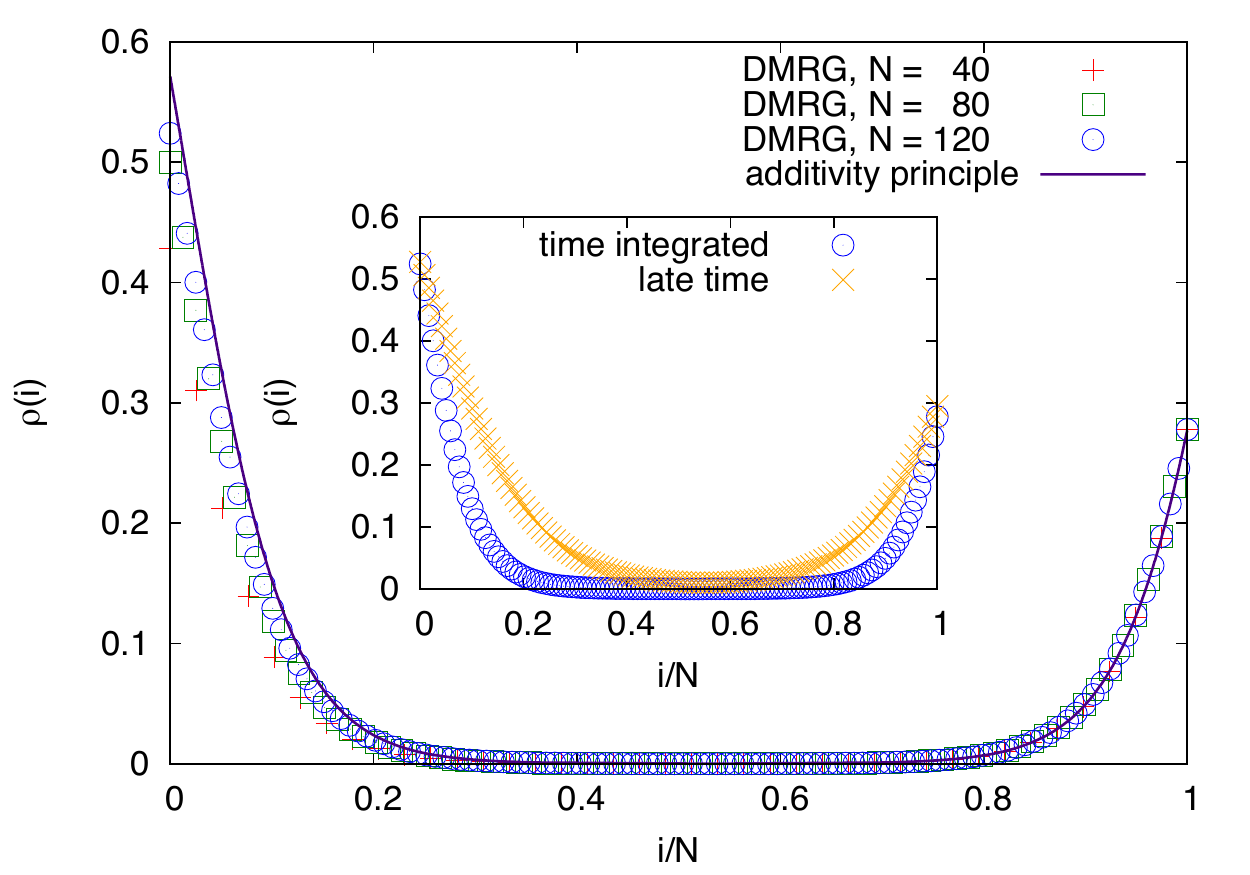}
\caption{\label{fig5} (Color online) Time-averaged density profile at $s=-10$ as calculated from the AP (full line) and from the DMRG. The inset shows a the time-averaged and the late time profile for $N=120$.}
\end{figure}

For various values of $s$ (or $j$) we have calculated density profiles in finite systems. We show as an example in Fig. \ref{fig4} the time-averaged profile corresponding to a large positive current fluctuation, $s=10$ (or $j=5.1214..$). The full line is the result from the AP, the symbols indicate DMRG data. As can be seen, the finite size results converge again to those predicted by the AP. This convergence is slowest near the boundaries. We therefore show in the inset an extrapolation of the average density at the leftmost side ($i=1$) for various $N$, which is consistent with the asymptotic prediction. The figure also shows that, apart from boundary effects, the density profile becomes flat and concentrated near $\rho=1/2$ in order to carry this current which is almost twice as large as the average one.
In Fig. \ref{fig5} we show similarly the density profile associated with a very small current, $s=-10, j=0.00041..$. Also for this profile with a minimum, finite $N$ results converge to the predictions of the AP. In this case the density becomes almost zero, except near the boundaries. In the inset we compare the time-averaged and late $T$-profile for a system with $N=120$. Late time profiles cannot be obtained from the AP. They are different from the time-averaged ones in the same way as spatial boundaries give rise to differences between bulk and surface densities \cite{Garrahan07}.

{\it Conclusions $-$} 
We have shown that current and density fluctuations in the WASEP are for $N$ sufficiently large precisely given by the AP. In contrast to the case on a ring \cite{Bodineau05} no evidence for a dynamic phase transition was found. This can be seen as a sort of non-equivalence between 'ensembles' since on a ring particle number is fixed  ('microcanonical') and for open boundaries it is not ('grand canonical'). It is well established that in equilibrium systems with long range interactions non-equivalence of ensembles can appear \cite{Touchette11}. Non-equilibrium systems like the WASEP have long range correlations {\it in time} \cite{Derrida07} and may therefore show similar phenomena. 

From a comparison between the asymptotic AP results and those from the DMRG it is possible to  quantify finite size corrections. This  can lead to a finite size scaling theory along the lines existing for the totally asymmetric exclusion process \cite{Gorissen09,Derrida98}. 

We have shown that the DMRG can give reliable results on density profiles in systems carrying a large fluctuation. It can therefore be used with confidence in non diffusive models or reaction-diffusion systems where so far few analytical approaches to large deviations exist. 

{\bf Acknowledgement} We would like to thank V. Lecomte for many useful discussions. We also thank J. Liesenborgs for help with numerical integration.


\begin{thebibliography}{100}
\bibitem{Gallavotti95}
G. Gallavotti and E.G.D. Cohen,  Phys. Rev. Lett. {\bf 74} 2694 (1995).
\bibitem{Lebowitz99}
J.L. Lebowitz and H. Spohn, J. Stat. Phys. {\bf 95} 333 (1999).
\bibitem{Bodineau04}
T. Bodineau and B. Derrida, Phys. Rev. Lett. {\bf 92} 180601 (2004).
\bibitem{Derrida07}
B. Derrida B, J. Stat. Mech.: Theory and Exp., P07023 (2007).
\bibitem{Hurtado09}
P.I. Hurtado and P.L. Garrido, Phys. Rev. Lett., {\bf 102}, 250601 (2009); P.I. Hurtado and P.L. Garrido, Phys. Rev. E {\bf 81}, 041102 (2010).  
\bibitem{Saito11}
K. Saito and A. Dhar, Phys. Rev. Lett., {\bf 107}, 250601 (2011). 
\bibitem{Bertini01}
L. Bertini, A. De Sole, D. Gabrielli, G. Jona-Lasinio and C. Landim, Phys. Rev. Lett. {\bf 87} 040601 (2001).
\bibitem{Bertini05}
L. Bertini, A. De Sole, D. Gabrielli, G. Jona-Lasinio and C. Landim, Phys. Rev. Lett. {\bf 94}, 030601 (2005). 
\bibitem{Bodineau05}
T. Bodineau and B. Derrida, Phys. Rev. E, {\bf 72}, 066110 (2005).
\bibitem{Hurtado11}
P.I. Hurtado and P.L. Garrido, Phys. Rev. Lett., {\bf 107}, 180601 (2011). 
\bibitem{JonaLasinio10}
G. Jona-Lasinio, Prog. Theor. Phys., {\bf 124} 731 (2010). 
\bibitem{Bodineau06}
T. Bodineau and B. Derrida, J. Stat. Phys., {\bf 123}, 277 (2006). 
\bibitem{White92}
S.R. White, Phys. Rev. Lett. {\bf 69} 2863 (1992).
\bibitem{Gorissen09}
M. Gorissen, J. Hooyberghs and C. Vanderzande, Phys. Rev. E {\bf 79} 020101(R) (2009); M. Gorissen and C. Vanderzande, J. Phys. A: Math. Theor. {\bf 44} 115005 (2011).
\bibitem{Spohn91}
H. Spohn, {\it Large Scale Dynamics of Interacting Particles} (Springer, Berlin, 1991).
\bibitem{Giardina06}
C. Giardin\`a, J. Kurchan and L. Peliti, Phys. Rev. Lett. {\bf 96}, 120603 (2006).
\bibitem{Tailleur07}
V. Lecomte and J. Tailleur, J. Stat. Mech.: Theory and Exp., P03004 (2007).
\bibitem{Gorissen11}
M. Gorissen, {\it Current fluctuations in exclusion processes: from code to codon} (Ph.D. thesis, Hasselt University, 2011).
\bibitem{Garrahan07}
J.P. Garrahan, R.L. Jack, V. Lecomte, E. Pitard, K. van Duijvendijk and F. van Wijland, J. Phys. A: Math. Theor., {\bf 42} 075007 (2009). 
\bibitem{Schollwock05}
U. Schollw\"{o}ck, Ann. Phys. {\bf 326}, 96 (2011).
\bibitem{Derrida93}
B. Derrida, M.R. Evans, V. Hakim and V. Pasquier,  J. Phys. A: Math. Gen. {\bf 26} 1493 (1993).
\bibitem{HurtadoPNAS11}
P.I. Hurtado, C. P\'erez-Espigares, J.J. del Pozo and P.L. Garrido, Proc. Natl. Acad. Sci U.S.A. {\bf 108}, 7704 (2011).  
\bibitem{Derrida98}
B. Derrida and J.L. Lebowitz, Phys. Rev. Lett. {\bf 80} 209 (1998); B. Derrida and C. Appert, J. Stat. Phys. {\bf 94} 1 (1999). 
\bibitem{Touchette11}
H. Touchette, Europhys. Lett. {\bf 96} 50010 (2011).


\end{thebibliography}
\end{document}